\documentclass[conference]{IEEEtran}
\IEEEoverridecommandlockouts

\usepackage[mode=buildmissing]{standalone} 
\usepackage{amsmath}
\usepackage{bm}
\usepackage[nolist]{acronym}
\usepackage{amssymb}
\usepackage{comment}
\usepackage{graphicx}
\usepackage{color,colortbl}
\usepackage[dvipsnames]{xcolor}
\usepackage{cleveref}
\usepackage{verbatim}
\usepackage{enumerate}
\usepackage[shortlabels]{enumitem}
\usepackage[boxed,ruled]{algorithm2e}

\usepackage{cuted, nccmath}
\usepackage{dblfloatfix}
\usepackage{float} 
\usepackage{lipsum}
\usepackage{balance}
\usepackage{cite}
\usepackage{url}
\usepackage{tabularx}
\usepackage{boldline}
\usepackage{balance}    
\usepackage{mathtools} 
\usepackage{stmaryrd} 
\usepackage{graphicx}
\usepackage{pgfplots}
\usepackage{booktabs}
\usepackage[font={footnotesize}]{caption}
\usepackage{subfig} 
\usepackage{tikz}
\usetikzlibrary{arrows.meta}
\tikzset{every picture/.style={line width=0.6pt}}
\usetikzlibrary{calc,positioning}
\usetikzlibrary{arrows.meta,
                backgrounds,
                chains,
                fit,
                quotes}

\newcommand{\bbold}{\bm{b}}
\newcommand{\complexset}[2]{ \mathbb{C}^{#1 \times #2}  }
\newcommand{\Ngrid}{N_{\rm{grid}}}
\newcommand{\xx}{\bm{x}}
\newcommand{\xxhat}{\widehat{\xx}}
\newcommand{\AAb}{\boldsymbol{A}}
\newcommand{\hermit}{\mathsf{H}}
\newcommand{\trpose}{\top}

\newcommand{\norm}[1]{\left\lVert#1\right\rVert}
\newcommand{\atx}{\bm{a}_{\text{tx}}}
\newcommand{\arx}{\bm{a}_{\text{rx}}}
\newcommand{\conj}{ {\ast} }

\newcommand{\yy}{ \bm{y} }
\newcommand{\yr}{ \bm{y}_r }
\newcommand{\yc}{ {y}_c }
\newcommand{\bmv}{ \bm{v} }

\newcommand{\bmA}{ \bm{A} }
\newcommand{\bmM}{ \bm{M} }
\newcommand{\bmb}{ \bm{b} }
\newcommand{\bmg}{ \bm{g} }
\newcommand{\bmtheta}{ \bm{\theta} }
\newcommand{\bmthetagrid}{ \bmtheta_{\text{grid}} }

\newcommand{\x}{ \bm{x} }
\newcommand{\m}{\bm{m}}
\newcommand{\llr}{\mathcal{L}}
\newcommand{\llrlog}{\mathcal{L}^{\rm{log}}}
\newcommand{\thetamax}{\theta_{\max}}
\newcommand{\thetamin}{\theta_{\min}}
\newcommand{\thetarange}{\bm{\theta}_{\text{range}}}
\newcommand{\varthetamax}{\vartheta_{\max}}
\newcommand{\varthetamin}{\vartheta_{\min}}
\newcommand{\thetamean}{\theta_{\text{mean}}}
\newcommand{\alphahat}{\widehat{\alpha}}
\newcommand{\thetahat}{\widehat{\theta}}

\newcommand{\pfa}{P_{\text{fa}}}
\newcommand{\pmd}{P_{\text{md}}}
\newcommand{\SNR}{\text{SNR}}
\newcommand{\dB}{\,\text{dB}}

\newcommand{\Jcomm}{\mathcal{J}_\text{comm}}
\newcommand{\Jangle}{\mathcal{J}_\text{angle}}
\newcommand{\Jisac}{\mathcal{J}_\text{ISAC}}

\newcommand{\hypz}{ \mathcal{H}_0 }
\newcommand{\hypone}{ \mathcal{H}_1 }
\newcommand{\hdet}{ \underset{\mathcal{H}_0}{\overset{\mathcal{H}_1}{\gtrless}} }
\newcommand{\etatilde}{\widetilde{\eta}}
\newcommand{\etabar}{\widebar{\eta}}

\newcommand{\etx}{ E_{\text{tx}} }

\newcommand{\degree}{^{\circ}}

\renewcommand{\Pr}{p}

\DeclarePairedDelimiter\abs{\lvert}{\rvert}%
\DeclarePairedDelimiter\absbigs{\big\lvert}{\big\rvert}%
\makeatletter
\newcommand*\rel@kern[1]{\kern#1\dimexpr\macc@kerna}
\newcommand*\widebar[1]{%
  \begingroup
  \def\mathaccent##1##2{%
    \rel@kern{0.8}%
    \overline{\rel@kern{-0.8}\macc@nucleus\rel@kern{0.2}}%
    \rel@kern{-0.2}%
  }%
  \macc@depth\@ne
  \let\math@bgroup\@empty \let\math@egroup\macc@set@skewchar
  \mathsurround\z@ \frozen@everymath{\mathgroup\macc@group\relax}%
  \macc@set@skewchar\relax
  \let\mathaccentV\macc@nested@a
  \macc@nested@a\relax111{#1}%
  \endgroup
}
\makeatother

\definecolor{mycolor}{RGB}{0,255,255}
\begin{acronym}[ACRONYM]
\acro{6G}{6th generation wireless systems}
\acro{AD}[AD]{autonomous driving}
\acro{AI}{artificial intelligence}
\acro{ADAS}{advanced driver-assistance system}
\acro{AE}{autoencoder}
\acro{AoA}{angle-of-arrival}
\acro{AoD}{angle-of-departure}
\acro{BCE}{binary cross-entropy}
\acro{CCE}{categorical cross-entropy}
\acro{CRB}{Cram\'{e}r-Rao bound}
\acro{CSI}{channel state information}
\acro{FMCW}{frequency modulated continuous wave}
\acro{IQ}{in-phase/quadrature} 
\acro{ISAC}{integrated sensing and communication}
\acro{JRC}{joint radar and communication}
\acro{MIMO}{multi-input multi-output}
\acro{MISO}{multi-input single-output}
\acro{MD-ML}{model-driven ML}
\acro{ML}{machine learning} 
\acro{MLE}{maximum likelihood estimation} 
\acro{MSE}{mean squared error}
\acro{NLL}{negative log-likelihood}
\acro{OFDM}{orthogonal frequency-division multiplexing}
\acro{PDF}{probability density function}
\acro{PSD}{power spectral density}
\acro{QPSK}{quadrature phase shift keying}
\acro{ReLU}{Rectified Linear Unit}
\acro{RMSE}{root mean squared error}
\acro{SER}{symbol error rate}
\acro{SNR}{signal-to-noise ratio}
\acro{ULA}{uniform linear array}
\end{acronym}

\begin{document}
\bstctlcite{IEEEexample:BSTcontrol}

\title{Model-Driven End-to-End Learning for Integrated Sensing and Communication
\thanks{This work was supported, in part, by a grant from the Chalmers AI Research Center Consortium (CHAIR), by the European Commission through the H2020 project Hexa-X (Grant Agreement no. 101015956) and by the MSCA-IF grant
888913 (OTFS-RADCOM). The work of C.~Häger was also supported by the Swedish Research Council under grant no. 2020-04718.}}
\author{Jos\'{e} Miguel Mateos-Ramos\IEEEauthorrefmark{1}, Christian H\"{a}ger\IEEEauthorrefmark{1}, \\Musa Furkan Keskin\IEEEauthorrefmark{1}, Luc Le Magoarou\IEEEauthorrefmark{2}, Henk Wymeersch\IEEEauthorrefmark{1}\\
\IEEEauthorrefmark{1}Department of Electrical Engineering, Chalmers University of Technology, Sweden \\
\IEEEauthorrefmark{2}Univ Rennes, INSA Rennes, CNRS, IETR-UMR 6164, Rennes, France
}
\maketitle

\begin{abstract}
\Ac{ISAC} is envisioned to be one of the pillars of 6G. However, 6G is also expected to be severely affected by hardware impairments. Under such impairments, standard model-based approaches might fail if they do not capture the underlying reality. To this end, data-driven methods are an alternative to deal with cases where imperfections cannot be easily modeled. 
In this paper, we propose a model-driven learning architecture for joint single-target \ac{MIMO} sensing and \ac{MISO} communication. We compare it with a standard neural network approach under complexity constraints. Results show that under hardware impairments, both learning methods yield better results than the model-based standard baseline. If complexity constraints are further introduced, model-driven learning outperforms the neural-network-based approach. Model-driven learning also shows better generalization performance for new unseen testing scenarios.
\end{abstract}
\IEEEoverridecommandlockouts
\IEEEpeerreviewmaketitle

\vspace{-3mm}
\section{Introduction} \label{sec:introduction}
Integrated sensing and communication (\Ac{ISAC}) has in the past few years become one of the key enabling technologies within the vision for 6G \cite{chiriyath2017radar,tan2021integrated,wymeersch2021integration}. In this 6G context, ISAC not only provides a means to reuse communication infrastructure for sensing purposes (either with dedicated or joint waveforms), it also provides a way to optimize the operation of the communication system itself, in the form of blockage prediction, radio mapping, and proactive resource allocation. 

\Ac{ISAC} can be broadly categorized as \emph{radar-centric} and \emph{communication-centric}. In radar-centric design, the aim is to provide communication capabilities on top of existing radar sensing infrastructure, as e.g., in 
\cite{lampel2019performance,lazaro2021car2car}. Generally, radar-centric designs exhibit poor communication performance, driven largely by hardware and cost constraints. On the other hand, communication-centric \ac{ISAC} relies on modifying communication waveforms and signal processing to enable high-resolution sensing. At a cost of a potential reduction in data rate, flexible sensing performance is attained, due to the high degree of freedom provided in communication signal optimization, including power allocation, beamforming design, and  scheduling \cite{liu2022integrated,zhang2021enabling}. 

Within  
communication-centric \ac{ISAC}, 
both the problem of signal design and that of signal processing have been traditionally treated under the umbrella of model-based signal processing. 
Model-based methods have important benefits, such as performance guarantees, explainability, and predictable computational complexity. However, they suffer from performance degradation under model mismatch and can be hard to derive when models are complicated. These issues have been addressed via data-driven designs, relying on \ac{ML} to solve the design problem or the signal processing problem (or both). 
Among \ac{ML}-based \ac{ISAC} works, we mention \cite{demirhan2022integrated,shao2022machine,chen2022joint,wu2021sensing,mu2021integrated,liu2022predictive,liu2022learning,mateos2022end}. In \cite{demirhan2022integrated}, the potential of \ac{ML} in ISAC is discussed.
A extensive survey on \ac{ML} in \ac{ISAC} is provided in \cite{shao2022machine}, though with an emphasis on sensing. In \cite{chen2022joint}, \ac{ML} is used in sensing to learn the model order. In \cite{wu2021sensing}, a two-level multi-task artificial neural network is proposed to replace the \ac{ISAC} receiver, which is shown to mitigate the imperfections and non-linearities of THz systems. In \cite{mu2021integrated}, \ac{ML}-aided beamforming in \ac{ISAC} is tackled, where a neural network learns the mapping from received signal to angle (see also \cite{liu2022transformer}). A vehicular beamforming scenario is considered in \cite{liu2022predictive,liu2022learning}, where \ac{ML} is applied to learn beamformers. Finally, in our previous work \cite{mateos2022end}, we developed an end-to-end approach for the \ac{ISAC} problem, applying an \ac{AE} \cite{Oshea2017} to account for hardware impairments. These ML-based approaches to \ac{ISAC} can operate under model mismatch and in general require little knowledge about the problem, other than the loss function. The drawbacks of this class of methods lies in the lack of performance guarantees, limited interpretability, and often high training complexity. 
\begin{figure}
    \centering
    \includegraphics[width=\linewidth]{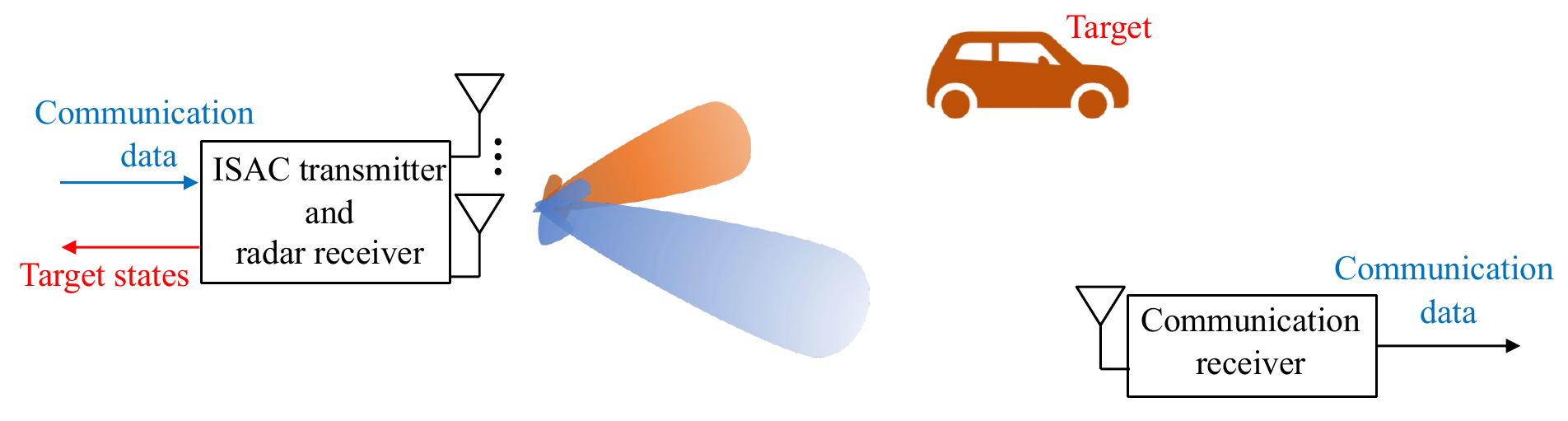}
    \caption{Considered \ac{ISAC} scenario with an \ac{ISAC} transmitter, co-located sensing/radar receiver, and a remote communication receiver. The learned transmit beams provide a flexible trade-off between sensing and communication.}
    \label{fig:SystemFigure}\vspace{-7mm}
\end{figure}

A third way, in between model-based and \ac{ML}-based \ac{ISAC} is provided by so-called \emph{\ac{MD-ML}} \cite{shlezingermodel}. This recent approach proved particularly relevant for communication systems \cite{Shlezinger2021, He2019}, due to the abundance of models characterizing them, as opposed to other domains of data processing such as image, sound or natural language. In \ac{MD-ML}, existing designs, algorithms, functional decompositions or protocols are used and reformulated as structured neural networks. An early example is the principle of deep unfolding \cite{Gregor2010,Hershey2014}, in which the iterations of an iterative method are interpreted as the layers of a neural network (see \cite{Balatsoukas2019} for a survey of deep unfolding applied to communication systems). 
As an added benefit, the neural network can be initialized from the model-based counterpart, thus starting from a already reasonably well performing point instead of at random. Especially relevant for ISAC are models linking sent or received signal at antenna arrays to directions of propagation. 
For instance, in \cite{yassine2022mpnet}, the steering vector models are taken as an initialization and made flexible by learning with a neural network of varying depth unfolding the matching pursuit algorithm \cite{Mallat1993}. The approach was then optimized and extended to the frequency response vector model (which has the same mathematical form) in \cite{Chatelier2022}. 

In this paper, we apply \ac{MD-ML} for end-to-end ISAC. Although \ac{MD-ML} has been investigated in communication scenarios \cite{Shlezinger2021, He2019, Balatsoukas2019, yassine2022mpnet, Chatelier2022}, there is no research on \ac{MD-ML} for sensing. Our main contribution is to apply \ac{MD-ML} to single-target sensing and extend this approach to obtain performance trade-offs between sensing and a \ac{MISO} communication link. 
As \ac{MD-ML} end-to-end learning, we optimize a model-based matrix of steering vectors to account for hardware impairments. 
However, unlike \cite{yassine2022mpnet}, we propose an architecture with parameter sharing to perform 2 tasks simultaneously: (i) precoder designing at the transmitter and (ii) target \ac{AoA} estimation at the receiver. 
A comparison between (i) \ac{MD-ML}, (ii) neural-network-based learning and (iii) the best-known baseline is made under hardware impairments and complexity constraints.

\section{System Model}
\label{sec:model}
    In this section, we describe the considered ISAC scenario, which is depicted in Fig.~\ref{fig:SystemFigure}. 
    
    \subsection{Sensing Model}
    \label{subsec:sensing_model}
        We consider a monostatic radar with a \ac{ULA} of $K$ antenna elements. At each transmission, the transmitter sends a complex signal $\bm{x} \in \mathbb{C}^K$, subject to $ \mathbb{E}[ \Vert \bm{x} \Vert^2 ] \le E_{\text{tx}}$. A single target in the environment might reflect the signal back to the transmitter. The probability that the target is present is drawn from a Bernoulli distribution $t \sim \text{Bern}(1/2)$. If a target is present ($t=1$), the received signal at the \ac{ULA} is
        \begin{align}
            \yr= 
            \alpha \bm{a}_{\text{rx}}(\theta)\bm{a}_{\text{tx}}^\top(\theta) \bm{x}+\bm{n}, 
            \label{eq:radar_scenario}
        \end{align}
        where we assume a Swerling-1 model of the target, such that $\alpha \sim \mathcal{CN}(0,\sigma_r^2)$, with $\sigma_r^2$ representing the loss of power due to path loss and the target's radar cross section. The steering vector is $[\bm{a}_{\text{tx}}(\theta)]_k=[\bm{a}_{\text{rx}}(\theta)]_k= \exp(-\jmath 2 \pi (k-(K-1)/2) d \sin (\theta ) / \lambda)$, with $d = \lambda / 2$ and $\lambda$ the wavelength. The \ac{AoA} (or \ac{AoD}) of the target $\theta$, is uniformly distributed as $\theta \sim \mathcal{U}[\thetamin, \thetamax]$. The prior knowledge $\{\thetamin, \thetamax\}$ is assumed to be available, with $-\pi/2 \leq \theta_{\min} \leq \theta_{\max} \leq \pi/2$. Regardless of the target presence, complex Gaussian noise $\bm{n} \sim \mathcal{CN}(\bm{0}, N_0\bm{I}_K)$ is added at the receiver side.
        
        The goal of the co-located receiver is to maximize the detection probability of the target, subject to some false alarm probability, and to estimate $\theta$ in the case of target detection, based on $\yr$.
        
    \subsection{Communication Model}
    \label{subsec:comm_model}
        We consider the same transmitter of $K$ antenna elements. The transmitter maps a message $m \in \mathcal{M}$ from a set of possible messages $\mathcal{M}$ into a complex symbol $s(m) \in \mathbb{C}$. The symbol $s(m)$ is precoded by $\bm{v} \in \mathbb{C}^K$ to steer the \ac{ULA} energy towards the receiver direction. The output signal is $\bm{x}(m) = \bm{v} s(m)$, again subject to $ \mathbb{E}[ \Vert \bm{x} \Vert^2 ] \le E_{\text{tx}}$. The receiver has a single antenna element, which yields a \ac{MISO} communication link. The communication receiver is always present, and the received signal follows the model
        \begin{align}
            \yc=\beta\bm{a}_{\text{tx}}^\top(\vartheta) \bm{x}(m)+{n},
            \label{eq:comm_scenario}
        \end{align}
        where a Rayleigh channel is considered, with $\beta  \sim \mathcal{CN}(0,\sigma_{c}^2)$ and $n \sim \mathcal{CN}(0, N_0)$, and the communication receiver is randomly located in a certain \ac{AoD} range $\vartheta \sim \mathcal{U}[\vartheta_{\min},\vartheta_{\max}]$, with prior knowledge of $\{\vartheta_{\min},\vartheta_{\max}\}$ and $-\pi/2 \leq \vartheta_{\min} \leq \vartheta_{\max} \leq \pi/2$. 
        We also assume that the receiver has access to the \ac{CSI} $\kappa = \beta\bm{a}_{\text{tx}}^\top(\vartheta) \bm{v}$ through a pilot sequence transmission. 
        
        The goal of the remote receiver is to retrieve the transmitted message based on the observation $\yc$.
        
    \subsection{Integrated Sensing and Communication}
    \label{subsec:isac_description}
        The purpose of \ac{ISAC} is to combine the sensing and communication transmitters into a joint transmitter that can be optimized to allocate energy between the target and the communication receiver direction. The transmitter considers the joint a priori angular information $\bm{\Theta} = \{\thetamin, \thetamax, \varthetamin, \varthetamax\}$. The receivers and the transmitter can be jointly optimized to obtain a desired trade-off between the communication and sensing performance.
        
\section{Baseline Approach}
\label{sec:baseline}
    The proposed learning approach is highly driven by the structure of standard model-based methods. Here we provide the derivation of the baseline, which is compared later with end-to-end learning approaches in Section \ref{sec:results}.
    
\subsection{Transmitter Benchmark}    
We design the benchmark for the transmit beamformer $\bm{x}$ in \eqref{eq:radar_scenario} or \eqref{eq:comm_scenario} by resorting to the beampattern synthesis approach in \cite{precoding_mmWave_JSTSP_2014,analogBeamformerDesign_TSP_2017}. To that end, we define a uniform angular grid covering $[-\pi/2, \pi/2]$ with $\Ngrid$ grid locations $\{\theta_i\}_{i=1}^{\Ngrid}$. For a given angular range $ \bm{\theta}_{\text{range}}$, 
 which could correspond to the direction of either the radar target or the communication receiver, we denote by $\bbold \in \complexset{\Ngrid}{1}$ the desired beampattern over the defined angular grid, given by
\begin{align}
    [\bbold]_i = 
    \begin{cases}
        K, &~~ \text{if} ~~\theta_i \in \bm{\theta}_{\text{range}} \\
        0, &~~ \text{otherwise.} 
    \end{cases} \label{eq:beampatternDesign2}
\end{align}
The problem of beampattern synthesis can then be formulated as 
$	\mathop{\mathrm{min}}\limits_{\xx} \norm{ \bbold - \AAb^\trpose \xx  }_{2}^2$, where $\AAb = \left[ \atx(\theta_1) \, \ldots \, \atx(\theta_{\Ngrid}) \right] \in \complexset{K}{\Ngrid}$ denotes the transmit steering matrix evaluated at the grid locations. This least-squares (LS) problem
has a simple closed-form solution 
\begin{equation}\label{eq:y_bench}
    \xx = (\AAb^\conj \AAb^\trpose)^{-1} \AAb^\conj \bbold,
\end{equation}
which yields, after normalization according to the transmit power constraints, a communication-optimal beam $\bm{x}_c$ or a radar-optimal beam $\bm{x}_r$.

For ISAC scenarios, a radar-communication trade-off beam is needed to provide adjustable trade-offs between the two functionalities. Using the approach from \cite{zhang2018multibeam}, we design the ISAC trade-off beam as
\begin{align} \label{eq:isac_precoder_bench}
    \bm{v}(\rho,\varphi) = \sqrt{E_{\text{tx}}}\frac{\sqrt{\rho} \bm{x}_r + \sqrt{1-\rho}e^{\jmath \varphi }\bm{x}_c}{\Vert\sqrt{\rho} \bm{x}_r + \sqrt{1-\rho}e^{\jmath \varphi }\bm{x}_c \Vert } ~,
\end{align}
where $\rho \in [0,1]$ is the ISAC trade-off parameter and $\varphi \in [0,2 \pi)$ is a phase ensuring coherency between multiple beams. 
By sweeping over $\rho$, we explore the ISAC performance of the baseline.

\subsection{Radar Detection Benchmark}    
Since the radar detection problem in \eqref{eq:radar_scenario} involves random parameters $\alpha$ and $\theta$, we derive the maximum a-posteriori (MAP) ratio test (MAPRT) detector \cite{MAP_Detector_TSP_2021} as our detector benchmark, which takes into account the prior information on $\alpha$ and $\theta$. Let $\hypz$ and $\hypone$ denote the absence and the presence of a target, respectively, in \eqref{eq:radar_scenario}. Then, the corresponding MAPRT is given by \cite{MAP_Detector_TSP_2021}
\begin{align}\label{eq_maprt}
    \llr(\yr) = \frac{ \max_{\alpha, \theta, \xx} p(\alpha, \theta, \xx, \hypone \, \lvert \, \yr ) }{ p(\hypz \, \lvert \, \yr ) } \hdet \etatilde ~.
\end{align}
Assuming $p(\hypz) = p(\hypone) = 1/2$, we find, after some derivation, that the test simplifies to\footnote{Although $\xx$ is known to the radar receiver, taking it as unknown in the MAPRT formulation \eqref{eq_maprt} and plugging in its optimal value as a function of $\theta$ simplify the detection test in \cite[Eq.~(16)]{mateos2022end} to a simple matched filter receiver in \eqref{eq_maprt_etabar_no_y}. This facilitates both the benchmark implementation in \eqref{eq_maprt_etabar_no_y} and the model-driven receiver design in Sec.~\ref{subsec:sensing_rx_training}.} (see \cite[App.~A]{mateos2022end})
\begin{align}\label{eq_maprt_etabar_no_y}
    \absbigs{ \arx^\hermit(\thetahat) \yr }^2   \hdet \etabar ~,
\end{align}
where $\etabar$ is a threshold determined based on a given false alarm probability and 
\begin{align}\label{eq_angle_est}
    \thetahat = \arg \max_{\theta \in [\thetamin, \thetamax]} \absbigs{ \arx^\hermit(\theta) \yr }^2    ~.
\end{align}
\subsection{Communication Receiver Benchmark}
\label{subsec:comm_rx_bench}
    Given the \ac{CSI} $\kappa = \beta \arx^\top(\theta)\bmv$, the received signal is $\yc = \kappa s(m) + n$. Hence, \ac{SER} is minimized by using \ac{MLE} as 
        $\hat{m}(\yc) = \arg\min_{m \in \mathcal{M}}\norm{ \yc - \kappa s(m)} ^2$.

\section{ISAC end-to-end Learning}
\label{sec:method}
\begin{figure*}[ht]
        \centering
        \begin{tikzpicture}[font=\scriptsize, >=stealth,nd/.style={draw,circle,inner sep=0pt,minimum size=5pt}, blkCH/.style={draw,minimum height=0.8cm,text width=1.9cm, text centered},  blk/.style={draw,minimum height=0.8cm,text width=2.2cm, text centered}, rounded/.style={circle,draw,minimum size=0.6cm, text centered},  x=0.6cm,y=0.55cm]
\tikzset{mult_fig/.pic={
\draw (-1,-1)--(1,1);
\draw (-1,1)--(1,-1);
}}
\tikzset{circ_fig/.pic={
\draw circle (1);
}}
\tikzset{circ_fill_fig/.pic={
\draw[fill=black] circle (1);
}}

\path		
	(-4,1.5)coordinate[](q){} 
	(-4,-1.5)coordinate[](theta_radar){} 
	node(Encoder)[blk, right=1.75 of q]{Encoder ($M$-QAM)}
	node(Beamformer_radar)[blk, right=1.75 of theta_radar]{$\bm{b} \in \{0,1\}^{\Ngrid}$}
	node(mult_M_radar) [rounded, right=0.5 of Beamformer_radar]{}
	node(M) [blk, text width = 0.8cm, minimum height = 0.5cm, below = 0.75 of mult_M_radar]{$\bmM(\bmA)$}
	node(mult_M_comm) [rounded, below=0.75 of M]{}
	node(Beamformer_comm)[blk, left=0.5 of mult_M_comm]{$\bm{b} \in \{0,1\}^{\Ngrid}$}
	coordinate[left=1.75 of Beamformer_comm](theta_comm){} 
	node(Combiner)[blk, text width=1cm, minimum height = 2.5cm, right=0.75 of M]{Joint ISAC precoder}
	node(mult) [rounded, right=4 of $(Encoder)!0.5!(Combiner)$]{}
    	node[blk,rounded corners,fill opacity=0, dashed, line width=0.5, minimum width=6.1cm, minimum height=4.8cm](Tx)at (2.65,-2){}
    	node[](Tx) at(6.5,-6){Transmitter}  
	coordinate[right=1.75 of mult](tx_out){} 
	node(Channel)[blkCH, below right=2 and 0.75 of tx_out]{Comm. channel \\ $\yc \sim p(\yc|\bm{x})$}
	node(Receiver)[blk, right=3 of Channel]{Comm.~Receiver (\ac{MLE})}
	coordinate[right=2.9 of Receiver](q_hat){} 
	coordinate[](right_bf) at (Combiner -| mult){} 
	coordinate[below=2 of right_bf](below_mult){} 
	coordinate[](below_ch) at (below_mult -| Channel){} 
	coordinate[](below_rec) at (below_mult -| Receiver){} 
	node(CH_radar)[blkCH, above right=1 and 0.75 of tx_out]{Radar channel \\ $\yr \sim p_t(\yr | \bm{x})$}
	node(mult_A) [rounded, right=2 of CH_radar]{}
	coordinate[above=1 of mult_A](mult_A_coor){}
	node(abs) [rounded, inner sep = -1, right=0.5 of mult_A]{$|\cdot|$}
	node(dot_mult_b) [rounded, below=1 of abs]{}
	coordinate[left=1 of dot_mult_b](dot_mult_b_coor){}
	node(softmax)[blk, text width=1cm, right=0.5 of dot_mult_b] {Softmax}	
	node(mult_theta) [rounded, right=0.75 of softmax]{}
	coordinate[above=1 of mult_theta](mult_theta_coor){}
	coordinate[right=2.3 of mult_theta](theta_hat){} 
;
\draw[->] (q)--node[above]{$m \in \mathcal{M}$}(Encoder);
\draw[->] (theta_radar)--node[above]{$\bm{\theta} \in \mathbb{R}^2$}(Beamformer_radar);
\draw[->] (theta_comm)--node[above]{$\bm{\vartheta} \in \mathbb{R}^2$}(Beamformer_comm);
\draw[->] (M)--(mult_M_radar);
\draw[->] (M)--(mult_M_comm);
\draw[->] (Encoder)-|node[above,pos=0.25]{$s(m) \in \mathbb{C}$}(mult);
\draw[->] (Beamformer_radar)--node[above]{}(mult_M_radar);
\draw[->] (mult_M_radar)--(mult_M_radar -| Combiner.west)node[above, pos=0.25]{$\bm{v}_r$};
\draw[->] (mult_M_comm)--(mult_M_comm -| Combiner.west)node[above, pos=0.25]{$\bm{v}_c$};
\draw[->] (Combiner)-|(mult)node[above, pos=0.25]{$\bm{v}$};
\draw[->] (Beamformer_comm)--node[above]{}(mult_M_comm);
\pic[scale=0.2](S) at (mult_M_radar) {mult_fig};
\pic[scale=0.2](S) at (mult_M_comm) {mult_fig};
\pic[scale=0.2](S) at (mult) {mult_fig};
\draw[-] (mult)--node[above]{$\bm{x} \in \mathbb{C}^K$}(tx_out);

\draw[->] (tx_out)|-(Channel);
\draw[->] (Channel)--node[above]{$\yc \in \mathbb{C}$}(Receiver);
\draw[->] (Receiver)--node[above]{$\hat{\m} \in [0,1]^{|\mathcal{M}|}$}(q_hat);
\draw[dotted] (right_bf)--(below_mult);
\draw[dotted] (Channel)--(below_ch);
\draw[dotted] (below_mult)--(below_ch);f
\draw[dotted] (below_ch)--node[above]{$\kappa = \beta \bm{a}_{\text{tx}}^\top(\vartheta)\bm{v}$}(below_rec);
\draw[dotted, ->] (below_rec)--(Receiver);

\draw[->] (tx_out)|-(CH_radar);
\draw[->] (CH_radar)--node[above]{$\yr \in \mathbb{C}^K$}(mult_A);
\draw[->] (mult_A)--node[]{}(abs);
\draw[->] (mult_A_coor)--node[above, pos = 0.1]{$\bmA^\hermit$}(mult_A);
\draw[->] (abs)--node[left]{$|\bmA^\hermit \yr|$}(dot_mult_b);
\draw[->] (dot_mult_b_coor)--node[left, pos = 0.1]{$\bmb$}(dot_mult_b);
\pic[scale=0.2](S) at (mult_A) {mult_fig};
    \pic[scale=0.27](S) at (dot_mult_b) {circ_fig};
    \pic[scale=0.05](S) at (dot_mult_b) {circ_fill_fig};
\draw[->] (dot_mult_b)--node[]{}(softmax);
\draw[->] (softmax)--node[above]{$\bmg$}(mult_theta);
\pic[scale=0.2](S) at (mult_theta) {mult_fig};
\draw[->] (mult_theta_coor)--node[above, pos = 0.1]{$\bmthetagrid$}(mult_theta);
\draw[->] (mult_theta)--node[above]{$\hat{\theta} = \bmg^\top \bmthetagrid$}(theta_hat);
\end{tikzpicture}
        \caption{Block diagram of the \ac{ISAC} model-driven approach. The matrix $\bmA \in \mathbb{C}^{K\times \Ngrid}$ is optimized via end-to-end learning by only considering single-target sensing. The communication encoder and receiver are implemented as in the baseline with no learnable parameters.}
        \label{fig:jrc_blocks} \vspace{-4mm}
    \end{figure*}
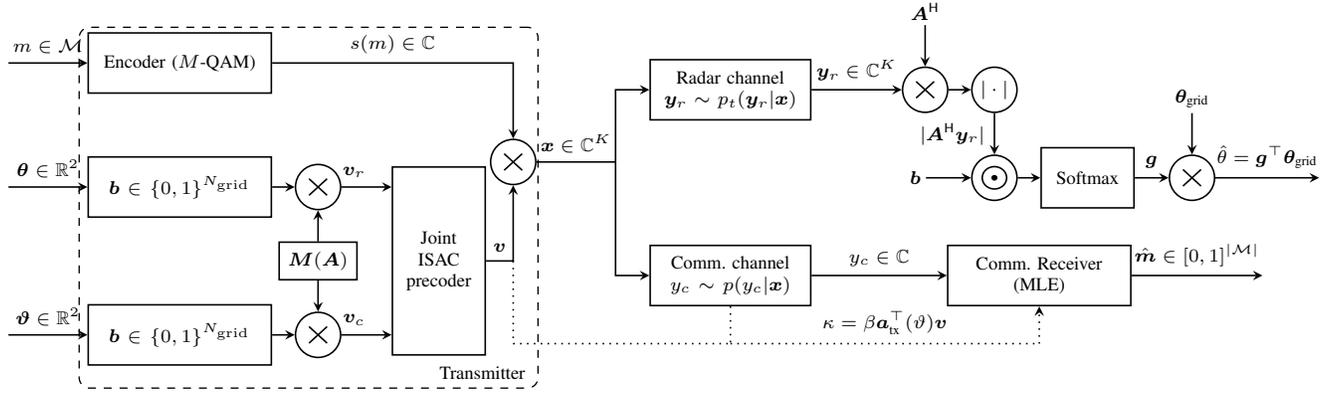

    In the following, we first describe the architecture of neural-network-based learning and the loss functions involved during training for the \ac{ISAC} scenario. Then, we specify how model-driven learning is trained, and how the \ac{ISAC} trade-offs are assessed with this approach. In Fig.~\ref{fig:jrc_blocks} we represent how the different components of the system are related for model-driven learning.
    
    \subsection{Neural-Network-Based End-to-End Learning}
    \label{subsec:unstructured_learning}
        We use 5 feed-forward neural networks as depicted in Table~\ref{tab:nn_architecture}. This \ac{AE} architecture is based on our previous work \cite{mateos2022end}, with some modifications that are stated in the following. On the transmitter side, the encoder $f_{\bm{\varepsilon}}: \{0,1\}^{|\mathcal{M}|} \to \mathbb{C}$ takes a one-hot encoded message $m$ and outputs a complex number interpreted as the symbol of a constellation. The beamformer $f_{\bm{\mu}}: \mathbb{R}^4 \to \mathbb{C}^K$ uses as input the prior information $\bm{\Theta}$ to yield a complex precoder $\bmv$. The radar receiver is divided into 2 networks: (i) for target detection, the network $f_{\bm{\rho}}: \mathbb{C}^{K+2} \to [0,1]$ concatenates the received signal $\yr$ and the sensing angular information $\{\thetamin, \thetamax\}$ as input, to predict the probability of the target in the environment as output; (ii) the angle estimator $f_{\bm{\nu}}: \mathbb{C}^{K+2} \to [-\pi/2, \pi/2]$ uses the same input as the presence detector, but it gives an estimate of the target angle $\thetahat$. The communication receiver $f_{\bm{\eta}}: \mathbb{C}^2 \to [0,1]^{|\mathcal{M}|}$ maps the concatenation of the received signal $\yc$ and the \ac{CSI} $\kappa$ to a vector of probabilities $\hat{\bm{m}}$, where the $i$-th element of $\hat{\bm{m}}$ represents the probability that the $i$-th message was transmitted. For complex-valued inputs or outputs, the concatenation of the real and imaginary parts is utilized. In Table~\ref{tab:nn_architecture}, the dimensions correspond to the real-valued concatenated vectors. The learnable parameters of each network are represented by $\bm{\varepsilon}, \bm{\mu}, \bm{\rho}, \bm{\nu}, \bm{\eta}$, respectively.
        
        \begin{table}[t]
        \centering
        \caption{Neural network architectures.}
        \label{tab:nn_architecture}
        \begin{tabular}{@{}cccc@{}}
        \toprule
        Network                & Input layer     & Hidden layers   & Output layer      \\ \midrule
        Encoder $f_{\bm{\varepsilon}}$ & $|\mathcal{M}|$ & $(K, K, 2K)$   & 2 (linear) \\
        Beamformer $f_{\bm{\mu}}$         & 4               & $(N, N, N)$   & $2K$ (linear) \\
        Presence det.~$f_{\bm{\rho}}$        & $2K+2$           & $(N, N, N)$ & 1 (sigmoid)       \\
        Angle est.~$f_{\bm{\nu}}$         & $2K+2$           & $(N, N, N)$ & 1 (tanh)       \\
        Comm.~receiver $f_{\bm{\eta}}$        & $4$           & $(K, 2K, 2K)$ & $|\mathcal{M}|$ (softmax)  \\ \bottomrule
        \end{tabular}
        \end{table}
        
        \subsubsection{Loss Functions}
        \label{subsec:loss_function}
            We choose a suitable loss function based on the task that needs to be solved.
            \begin{itemize}
                \item \textbf{Target Detection.} We use the \ac{BCE} loss. Let $\hat{p} \in [0,1]$ be the estimate of the probability that the target is in the environment. The \ac{BCE} loss is then
                \begin{equation}
                    \mathcal{J}_{\text{TD}} = -\mathbb{E}[t\log(\hat{p}) + (1-t)\log(1-\hat{p})],
                    \label{eq:radar_loss}
                \end{equation}
                
                \item \textbf{Angle Estimation.} We use the \ac{MSE} loss function between the estimated and true angles according to $\Jangle = \mathbb{E}[|\hat{\theta} - \theta|^2]$. Note that this loss is meaningful only when $t = 1$ and the neural network estimates that there is a target. We compute the \ac{MSE} loss in degrees.
                
                \item \textbf{Communication Message Estimation.} We treat this problem as a multi-class classification problem, where each of the classes corresponds to a transmitted message. Hence, we resort to the \ac{CCE} loss function. Let $\m^{\text{enc}} \in \{0,1\}^{|\mathcal{M}|}$ be the one-hot encoding of $m$ and $\hat{\m}\in [0,1]^{|\mathcal{M}|}$ a $|\mathcal{M}|$-dimensional probability vector. Then, the \ac{CCE} loss is 
                \begin{equation}
                    \Jcomm = -\mathbb{E} \left[ \sum_{j=1}^{|\mathcal{M}|} m^{\text{enc}}_j \log(\hat{m}_j) \right].
                \end{equation}
            \end{itemize}
                
        \subsubsection{Neural-network Based ISAC}
            To obtain a good \ac{ISAC} trade-off, we could train all 5 networks from Table \ref{tab:nn_architecture} at the same time. However, the scale of the detection and angle estimation loss functions might differ in several orders of magnitude. Then, we apply a 2-step learning procedure. Firstly, we learn $\bm{\varepsilon}, \bm{\mu}, \bm{\rho}, \bm{\eta}$ based on target detection using the following loss function
            \begin{equation} \label{eq:isac_loss_nn_pd}
                \Jisac^{\text{TD}} = \omega_r\mathbb{I}\{t = 1\}\mathcal{J}_{\text{TD}}  + (1-\omega_r)\Jcomm, 
            \end{equation}
            where $\omega_r \in [0,1]$ is a hyperparameter that allows for flexible trade-offs between sensing and communication performance, and $\mathbb{I}\{\cdot\}$ is the indicator function. Secondly, we learn $\bm{\nu}$ and update $\bm{\varepsilon}, \bm{\mu}, \bm{\eta}$ by training for angle estimation, using the joint loss function
            \begin{equation} \label{eq:isac_loss_nn_angle}
                \Jisac^{\text{angle}} = \omega_r\mathcal{J}_{\text{angle}}  + (1-\omega_r)\Jcomm. 
            \end{equation}
    
    \subsection{Model-Driven end-to-end Learning}
        
        \subsubsection{Trainable Model-Driven Transmitter}
        \label{subsec:transmitter_learning}
            According to the benchmark, the transmitter precoder is based on \eqref{eq:y_bench}, which involves the vector $\bmb$ and the matrix of steering vectors $\bmA$. The binary vector $\bmb$ is completely determined by the prior angular information $\thetarange$. For joint training, $\thetarange$ accounts for the target and the communication receiver at the same time. We let $\bmA \in \mathbb{C}^{K\times \Ngrid}$ be a matrix of complex trainable parameters. In this way, the matrix $\bmA$ is able to adapt to the hardware impairments in the transmitter \ac{ULA} that are described in Section \ref{subsec:results_hwi}.
            
            Nevertheless, \eqref{eq:y_bench} involves a matrix inversion which could yield numerical instability for not well-defined matrices $\bmA$ during training. Therefore, we instead compute a matrix $\bmM$, which is the result of solving the linear matrix equation $(\bmA^* \bmA^\top)\bmM = \bmA^*$, so $\bmM$ can be expressed in terms of $\bmA$ as
            \begin{equation}
                \bmM (\bmA) = (\bmA^* \bmA^\top)^+ \bmA^* ~,
            \end{equation}
            where $^+$ denotes the Moore-Penrose inverse. Then, the transmitter signal is simply $\x = \bmM \bmb$, which is then normalized to have energy $E_{\text{tx}}$.
            
        \subsubsection{Trainable Model-Driven Sensing Receiver}
        \label{subsec:sensing_rx_training}
            The test statistic in \eqref{eq_maprt_etabar_no_y} to compute the probability of detection is based on the angle estimation of the target. Hence, we imitate the same kind of procedure during learning, i.e., we only train $\bmA$ to yield a good angle estimate $\hat{\theta}$. 
            Note that the same matrix $\bmA$ is shared between transmitter and receiver. Moreover, the angle estimation from the benchmark in \eqref{eq_angle_est} resorts to finding the argument that maximizes the test statistic. This operation is not differentiable, and we compute instead  
            \begin{equation}
                \bmg = \text{softmax}(|\bmA^\hermit \yr| \odot \bmb),
            \end{equation}
            where we first compute the test statistic $|\bmA^H\yr|$ similarly to the benchmark, but we restrict this metric to be within $\thetarange$ by means of the element-wise product with $\bmb$.
            Ideally, $\bmg$ should be close to 1 in the position corresponding to the true angle (recall that in the ideal case, the columns of $\bmA$ represent steering vectors at different angles). Hence, by computing $\bmg^\top \bmthetagrid$, we expect to obtain a close estimation of the true angle. 
            
            Regarding target detection, even though we do not train the system for this task, we mimic \eqref{eq_maprt_etabar_no_y} and perform detection based on
            \begin{equation}
                \max\{ |\bmA^\hermit \yr| \odot \bmb \} \hdet \tilde{\eta},
            \end{equation}
            for some threshold $\tilde{\eta}$.

        \subsubsection{Model-based Communication Components}
        \label{subsec:comm_rx_learning} 
            We use a standard $|\mathcal{M}|$-QAM encoder for the transmitter and the \ac{MLE} approach from Section \ref{subsec:comm_rx_bench} at the receiver, which is optimal given the \ac{CSI}. However, no parameters are trained for the communication link.

        \subsubsection{Model-driven ISAC}
        \label{subsec:mdriven_isac}
            We note that, in contrast to neural-network-based learning, there is no need to directly train for \ac{ISAC}, since the communication encoder and receiver are implemented following the baseline. Moreover, once we train $\bmA$ for sensing purposes, the transmitter can be used to point towards different directions (given different inputs). Hence, we train $\bmA$ solely for single-target sensing. After that, to evaluate the \ac{ISAC} trade-offs, we create a joint precoder based on \eqref{eq:isac_precoder_bench}, with $\rho = \omega_r$ and $\varphi = 0$. In Fig.~\ref{fig:jrc_blocks}, it is depicted how we use different inputs to create a radar precoder ($\bmv_r$) and a communication precoder ($\bmv_r$), which are combined later following \eqref{eq:isac_precoder_bench} to yield the \ac{ISAC} precoder $\bmv \in \mathbb{C}^K$.

\section{Results}
\label{sec:results}
    In this section, we compare the performance of (i) model-driven learning, (ii) neural-network-based learning, and (iii) the model-based baseline described in Section \ref{sec:baseline}. 
    
    \subsection{Parameter Selection, Random Training, and Evaluation}
    \label{subsec:parameters_training}
        On the transmitter side, we consider an \ac{ULA} with $K = 16$ antenna elements, $E_{\text{tx}}=1$, and $|\mathcal{M}| = 4$ possible messages, which corresponds to a \ac{QPSK} constellation in the baseline approach. 
        The average radar \ac{SNR} is chosen as $\SNR_r = \sigma_r^2/N_0 = 0\dB$ and the average communication \ac{SNR} as $\SNR_c = \sigma_c^2/N_0 = 20\dB$.
        
        For simplicity, we assume that the communication receiver is located at a random position within a fixed angular sector $[\varthetamin, \varthetamax] = [30\degree, 50\degree]$. However, in the sensing scenario, we randomize the angular sector of the target as in \cite{rivetti2022spatial}. We first draw the mean angle of the sector as $\thetamean \sim \mathcal{U}[-60\degree, 60\degree]$ and the span as $\Delta = \mathcal{U}[10\degree, 20\degree]$. The target prior information is then $\{\thetamin, \thetamax\} = \{\thetamean - \Delta/2, \thetamean + \Delta/2\}$. However, we show only results corresponding to a testing interval of $[\thetamin, \thetamax] = [-40\degree, -20\degree]$.
        
        We use the Adam optimizer \cite{kingma2015adam} for the learning approaches, with a learning rate of $10^{-3}$ and a batch size of 10,000 samples. 
        In model-driven learning, the matrix $\bmA$ is initialized as a perturbed version of the baseline steering matrix, i.e., $[\bmA]_{m,l} = \exp(-\jmath \pi (m-(K-1)/2) \sin(\theta_l)) + n$, with $n \sim \mathcal{N}(0,0.1)$. The values for the trade-off parameter are $\omega_r \in \{0, 10^{-6}, 10^{-5}, 10^{-4}, 10^{-3}, 10^{-2}, 0.05, 0.1, 0.15, 0.4, 0.6, 0.8,\allowbreak 1\}$. For each $\omega_r$ value, we retrain all neural networks from scratch.
        
        During the testing stage, we evaluate the performance of each method by computing the probability of misdetection $\pmd = \Pr(\hat{t} = 0 | t = 1)$, the \ac{SER} $\Pr(\hat{m} \neq m)$, and the sensing angle \ac{RMSE} $\sqrt{E[|\hat{\theta} - \theta|^2]}$ for a given false alarm probability $\pfa = \Pr(\hat{t} = 1 | t = 0) = 10^{-2}$. The \ac{RMSE} is calculated only when the target is present and it has been detected by the receiver.
        
    \subsection{Results without Hardware Impairments}
    \label{subsec:results_ideal}
        We first consider the case of ideal conditions in the \ac{ULA} array ($d = \lambda/2$), without complexity restrictions. We set the number of hidden neurons in the neural networks of the sensing \ac{AE} as $N = 1024$, giving approximately 6.4 million real-valued trainable parameters. We also fixed a grid of $\Ngrid = 500$ discrete angles, resulting in  8000 complex-valued trainable parameters for the model-driven learning architecture. 
        Further increasing the number of parameters did not yield significant performance improvement.
        The number of training iterations is set to 50,000.
        
\begin{figure}[t]
        \captionsetup[subfigure]{labelformat=empty}
            \centering
            \subfloat[]{\begin{tikzpicture}[font=\scriptsize]
\begin{axis}[
width=8cm,
height=4.cm,
legend cell align={left},
legend style={
  fill opacity=1,
  draw opacity=1,
  text opacity=1,
  at={(0.6,0.4)},
  anchor=south west
},
log basis y={10},
tick align=outside,
tick pos=left,
x grid style={white!69.0196078431373!black},
xlabel={Misdetection probability},
xmajorgrids,
xmin=0, xmax=0.8,
xtick style={color=black},
y grid style={white!69.0196078431373!black},
ylabel={SER},
ymajorgrids,
ymin=1e-03, ymax=1,
ymode=log,
ytick style={color=black},
]
\addplot+[only marks, color=red, mark=triangle*, mark options={scale=0.8}]
coordinates { 
(0.990148465101674,0.0017146687080748096)
(0.9892214126052092,0.0017087678990143152)
(0.9903776423123788,0.0017567321511069195)
(0.9899917566078056,0.001552477234746523)
(0.9895024593563577,0.001862140796431128)
(0.9894564901283545,0.0018464162354304727)
(0.9890732611155598,0.001545533373526791)
(0.9895785486468961,0.0016920975275847494)
(0.3538556826051322,0.0020361920391215485)
(0.3578670216894525,0.0021539657468046)
(0.358837559862657,0.001925003673828764)
(0.3626423965476834,0.0018931768993627055)
(0.3661112665654951,0.0020900838305980547)
(0.3676216961315091,0.0020958142067571514)
(0.3622292158901247,0.0019068270823102953)
(0.3593942941238105,0.0019882837316949815)
(0.20797964309930395,0.00244)
(0.209235665832055,0.0022764629443900682)
(0.214148358355298,0.0024421999585789844)
(0.21299259272459714,0.0023913061588687326)
(0.2183382801293835,0.0022103649633219056)
(0.21294154523112774,0.002602303372161316)
(0.21565298842701597,0.002201019713902579)
(0.21114662199532097,0.002468567188233315)
(0.14967563046851717,0.002822289943019212)
(0.1500967740038548,0.0030709434006640835)
(0.15119415787936397,0.0030785836489153214)
(0.15146843843067792,0.002707194978737572)
(0.15275126212215973,0.002861081559510243)
(0.15426157089996506,0.0028474393736794227)
(0.15304323852459745,0.0028633666881691154)
(0.14953775585511497,0.0029406065010677794)
(0.11669989719360019,0.003779245103328335)
(0.11430302226154931,0.00394349040134301)
(0.11778393984174096,0.004061030706303125)
(0.1168286243291291,0.004300563278319885)
(0.11812915099865462,0.004088120059226488)
(0.11742796348364926,0.004097028674274423)
(0.1179445834816163,0.00378597046677008)
(0.11836920898304282,0.00385001537736738)
(0.09420015189871722,0.005286468554770532)
(0.09457710198953573,0.005926561737046926)
(0.09657623211791844,0.005616847691813659)
(0.0952042046536512,0.005720710812197031)
(0.09483729035138766,0.005884738964072892)
(0.09417608078056572,0.0060955432723621755)
(0.09511110101724285,0.00588)
(0.09570415704967206,0.006246551862234682)
(0.07988953126849518,0.011075237142086915)
(0.08093543098023004,0.011321941495506202)
(0.08138976584655944,0.010980164683304726)
(0.08199177579393924,0.011650781393243999)
(0.08243137415508273,0.01200968910837351)
(0.08140558446089086,0.01239736371732817)
(0.08000983794085048,0.012050419566129743)
(0.08029187699341289,0.01161947232103996)
(0.07034469314567349,0.5716874185244968)
(0.06864800082621858,0.5711641571235021)
(0.06903432078405103,0.5709699498955848)
(0.06995192203289602,0.5699361147101281)
(0.06927588487420155,0.5727006245264411)
(0.07040737245464035,0.5716943935102508)
(0.07077361840422147,0.5711048442946792)
(0.07054593163168354,0.5724526734972853)
};\addlegendentry{Baseline};
\addplot +[thick, color=blue!50!black, solid, mark=*, mark options={scale=0.8,solid }] 
coordinates { 
(0.9892614473262888,0.00131)
(0.9894582246923745,0.00129)
(0.9901478946525777,0.00129)
(0.9904609067450227,0.00137)
(0.8952578049951969,0.00163)
(0.25122198206852386,0.00193)
(0.13989071005443487,0.00319)
(0.12462696125520334,0.00364)
(0.11500204268662984,0.00421)
(0.09650131284021768,0.00731)
(0.08643055555555557,0.01254)
(0.08510725161976007,0.01784)
(0.08003554274735836,0.75069)
};
\addlegendentry{NN learning}
\addplot +[thick, color=green!60!black, solid, mark=diamond*, mark options={scale=0.8,solid }] 
coordinates { 
(0.9912740948728485,0.00155)
(0.5004574060015183,0.00147)
(0.3102191231746665,0.00176)
(0.22682024480170682,0.00199)
(0.18152219697645178,0.00222)
(0.14977751325103572,0.0025)
(0.12789312186179957,0.00328)
(0.11112794451335894,0.00373)
(0.09667717848116597,0.00467)
(0.08761937980356294,0.00648)
(0.07964885742202965,0.00988)
(0.07069734573593678,0.01845)
(0.06493407910507387,0.64175)
};
\addlegendentry{MD learning}

\end{axis}

\end{tikzpicture}

}
            \vspace{-4mm}
            \subfloat[]{\begin{tikzpicture}[font=\scriptsize]
\begin{axis}[
width=8cm,
height=4.cm,
legend cell align={left},
legend style={
  fill opacity=1,
  draw opacity=1,
  text opacity=1,
  at={(0.6,0.4)},
  anchor=south west
},
log basis y={10},
tick align=outside,
tick pos=left,
x grid style={white!69.0196078431373!black},
xlabel={RMSE [deg]},
xmajorgrids,
xmin=0, xmax=10,
xtick style={color=black},
y grid style={white!69.0196078431373!black},
ylabel={SER},
ymajorgrids,
ymin=1e-03, ymax=1,
ymode=log,
ytick style={color=black},
]
\addplot+[only marks, color=red, mark=triangle*, mark options={scale=0.8}]
coordinates { 
(8.69104820926478,0.0017146687080748096)
(9.10005139837629,0.0017087678990143152)
(8.439989187617776,0.0017567321511069195)
(8.603787194875304,0.001552477234746523)
(8.871676215483935,0.001862140796431128)
(8.904618055779103,0.0018464162354304727)
(8.593961784475635,0.001545533373526791)
(8.705796848027322,0.0016920975275847494)
(1.2269812522268289,0.0020361920391215485)
(1.223959653605897,0.0021539657468046)
(1.1809168453766274,0.001925003673828764)
(1.240873008853414,0.0018931768993627055)
(1.204970126147865,0.0020900838305980547)
(1.1972153900294096,0.0020958142067571514)
(1.2268603401623934,0.0019068270823102953)
(1.2380795473379522,0.0019882837316949815)
(0.9782919908598897,0.00244)
(0.9622765525550421,0.0022764629443900682)
(0.9861694711812936,0.0024421999585789844)
(0.9661038268574844,0.0023913061588687326)
(0.970618128110921,0.0022103649633219056)
(0.9683900718089826,0.002602303372161316)
(0.9501875314215348,0.002201019713902579)
(0.9695404171788512,0.002468567188233315)
(0.8465912582996915,0.002822289943019212)
(0.849193115217727,0.0030709434006640835)
(0.8593583479112535,0.0030785836489153214)
(0.8548544257571634,0.002707194978737572)
(0.8456666090942849,0.002861081559510243)
(0.8480335940184875,0.0028474393736794227)
(0.8334236916150208,0.0028633666881691154)
(0.8054091133301708,0.0029406065010677794)
(0.7752378049999817,0.003779245103328335)
(0.7986998768197788,0.00394349040134301)
(0.7741242095538561,0.004061030706303125)
(0.7817062196721217,0.004300563278319885)
(0.7579328188234462,0.004088120059226488)
(0.7799269296175965,0.004097028674274423)
(0.7446641790827725,0.00378597046677008)
(0.7553203350204578,0.00385001537736738)
(0.698006554206475,0.005286468554770532)
(0.7144145399583353,0.005926561737046926)
(0.71163626496386,0.005616847691813659)
(0.6913433879224792,0.005720710812197031)
(0.7150403293339344,0.005884738964072892)
(0.7362327587754244,0.0060955432723621755)
(0.7395616544473218,0.00588)
(0.7319305404454943,0.006246551862234682)
(0.6621872044928037,0.011075237142086915)
(0.6694811524460951,0.011321941495506202)
(0.6598818613830245,0.010980164683304726)
(0.6810259674806468,0.011650781393243999)
(0.6687122706473186,0.01200968910837351)
(0.6712836661081286,0.01239736371732817)
(0.6578401374354421,0.012050419566129743)
(0.6577006764021037,0.01161947232103996)
(0.6308686099230567,0.5716874185244968)
(0.6531785415508997,0.5711641571235021)
(0.661856839391274,0.5709699498955848)
(0.6333425745977466,0.5699361147101281)
(0.6420001286942816,0.5727006245264411)
(0.619717198015258,0.5716943935102508)
(0.619852664027562,0.5711048442946792)
(0.6084876644196623,0.5724526734972853)
};\addlegendentry{Baseline};
\addplot +[thick, color=blue!50!black, solid, mark=*, mark options={scale=0.8,solid }] 
coordinates { 
(26.577612807167327,0.00131)
(5.633057390403969,0.00129)
(5.715051240311619,0.00129)
(6.088470600835923,0.00137)
(2.526930123491256,0.00163)
(0.9600716102985868,0.00193)
(0.8250328500741673,0.00319)
(0.7288100476591944,0.00364)
(0.7256627186413067,0.00421)
(0.6724407711487541,0.00731)
(0.6542837613328033,0.01254)
(0.678033816332464,0.01784)
(0.6119156234570022,0.75069)
};
\addlegendentry{NN learning};
\addplot +[thick, color=green!60!black, solid, mark=diamond*, mark options={scale=0.8,solid}]
coordinates { 
(8.011421064218977,0.00155)
(1.5617671162386635,0.00147)
(1.2506254982408187,0.00176)
(1.0867773643573002,0.00199)
(0.9562474547811584,0.00222)
(0.8848618837603459,0.0025)
(0.8236592610925113,0.00328)
(0.8087890579868233,0.00373)
(0.7440878664811641,0.00467)
(0.7477334597396652,0.00648)
(0.7139221951149685,0.00988)
(0.6963911034133482,0.01845)
(0.7021523942251235,0.64175)
};
\addlegendentry{MD learning};

\end{axis}

\end{tikzpicture}

}
            \vspace{-4mm}
            \caption{Results without hardware impairments and without complexity restrictions. The target lies in the angular sector $[\thetamin, \thetamax] = [-40\degree, -20\degree]$. The communication receiver is randomly located in the angle interval $[\thetamin, \thetamax] = [30\degree, 50\degree]$. The \ac{ISAC} scenario has fixed testing parameters $\pfa = 10^{-2}$, $\SNR_c = 20\dB$, and $\SNR_r = 0\dB$.}
            \label{fig:ser_vs_pd_1} \vspace{-4mm}
        \end{figure}
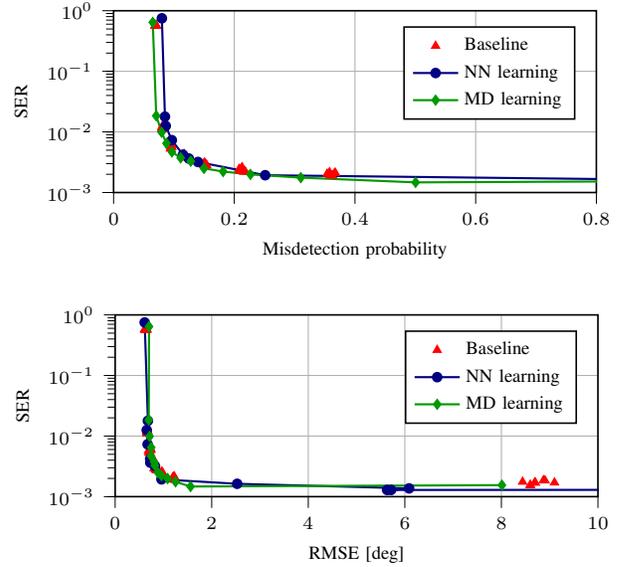

        Fig.~\ref{fig:ser_vs_pd_1} shows  the \ac{ISAC} results for one particular testing angular sector.
        No significant differences can be observed between the learning approaches and the baseline. 
        Indeed, for this scenario the baseline transmitter and receiver algorithms are either optimal (for communications) or close to optimal (for radar sensing).  
        Moreover, without complexity constraints both learning approaches can be trained to perform similar to the baseline. 
        
        
    \subsection{Results under Hardware Impairments}
    \label{subsec:results_hwi}
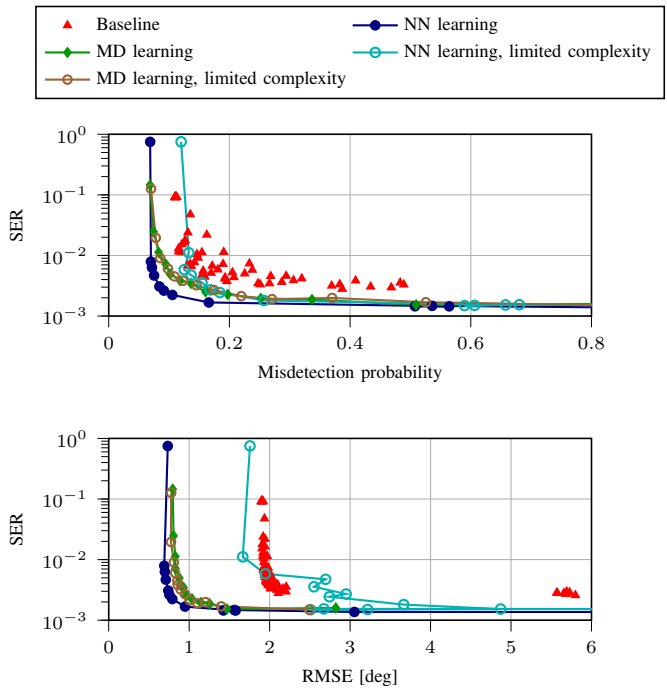
\begin{figure}[t]
        \captionsetup[subfigure]{labelformat=empty}
            \centering
            \subfloat[]{\begin{tikzpicture}[font=\scriptsize]
\begin{axis}[
width=8cm,
height=4.cm,
legend cell align={left},
legend columns = 2,
legend style={
  fill opacity=1,
  draw opacity=1,
  text opacity=1,
  at={(0.5,1.45)},
  anchor=center
},
log basis y={10},
tick align=outside,
tick pos=left,
x grid style={white!69.0196078431373!black},
xlabel={Misdetection probability},
xmajorgrids,
xmin=0, xmax=0.8,
xtick style={color=black},
y grid style={white!69.0196078431373!black},
ylabel={SER},
ymajorgrids,
ymin=1e-03, ymax=1,
ymode=log,
ytick style={color=black},
]
\addplot+[only marks, mark=triangle*, mark options={scale=0.8,color=red}]
coordinates { 
(0.9684216064150275,0.00296)
(0.9690959474274152,0.00283)
(0.9715461510834431,0.0028)
(0.9685908158976074,0.00258)
(0.9683444396822508,0.00272)
(0.97154740195454,0.00287)
(0.9696418446858558,0.0028)
(0.9691017499103121,0.00271)
(0.4091594352854614,0.00385)
(0.38280147288521993,0.00334)
(0.3694835388080917,0.00316)
(0.3869410034610179,0.0028)
(0.4340536993792786,0.00305)
(0.46829945442875476,0.00296)
(0.48903904550398847,0.0033)
(0.4830831786704983,0.00355)
(0.2940934793829213,0.00459)
(0.2684957640975858,0.00455)
(0.2475307948564789,0.00346)
(0.2511613886613886,0.00331)
(0.2664093027051151,0.00345)
(0.28682580972005056,0.00363)
(0.3061958082620506,0.00389)
(0.3199166004765688,0.00416)
(0.23314224034540654,0.00735)
(0.20504056853673802,0.00526)
(0.1928196252197374,0.00432)
(0.192883987319854,0.00411)
(0.19578866466812872,0.00376)
(0.20766777488483878,0.00436)
(0.22536503647811934,0.00502)
(0.23806544797546647,0.00578)
(0.19055245388722197,0.01136)
(0.1715549196741456,0.00674)
(0.15650444008764852,0.00558)
(0.15450552838523857,0.00473)
(0.1594672757874489,0.00445)
(0.1696704223101203,0.00515)
(0.18079959079959085,0.00584)
(0.18951209159799576,0.00722)
(0.162574239439087,0.02192)
(0.14617476504699067,0.01047)
(0.13503678594815305,0.00726)
(0.1308533335273776,0.00682)
(0.13699894892487485,0.0066)
(0.14211612434417975,0.0077)
(0.1482534944636824,0.00912)
(0.15445427659466404,0.01121)
(0.13536704642190867,0.04755)
(0.12521511413471842,0.01647)
(0.12024367240209188,0.01123)
(0.11599266226402516,0.01161)
(0.11572224971699341,0.01299)
(0.12133468004530878,0.01522)
(0.12680638668932276,0.0178)
(0.13130301144613532,0.02384)
(0.11283988642600673,0.09174)
(0.1097754242231942,0.09311)
(0.11002591716450827,0.09068)
(0.11133926216623946,0.09205)
(0.11050033250117197,0.0904)
(0.11185790359684245,0.09178)
(0.11054980833052164,0.09241)
(0.11186320811928885,0.09226)
};\addlegendentry{Baseline};
\addplot +[thick, color=blue!50!black, solid, mark=*, mark options={scale=0.8,solid }] 
coordinates { 
(0.9866461307983319,0.001366)
(0.9063308927000879,0.001367)
(0.5359693886238521,0.001473)
(0.5075439748075355,0.001447)
(0.5642430122852133,0.001445)
(0.16589108212608483,0.001676)
(0.10548666666666662,0.002231)
(0.09149155596153824,0.002624)
(0.08404670482568977,0.003079)
(0.07523874612478654,0.004664)
(0.07173465154595393,0.006301)
(0.06999019964024822,0.007911)
(0.06872249040537304,0.749921)
};
\addlegendentry{NN learning}
\addplot +[thick, color=green!60!black, solid, mark=diamond*, mark options={scale=0.8,solid }] 
coordinates { 
(0.8946221037072547,0.00159)
(0.5093530986534542,0.00153)
(0.3368406346671988,0.00187)
(0.2519815860642698,0.00195)
(0.1971248312820718,0.00227)
(0.1605535202461289,0.00251)
(0.13638559018091967,0.00346)
(0.11841084790110601,0.00383)
(0.10160951237088633,0.00495)
(0.09462514047955062,0.00716)
(0.08278209458707719,0.01127)
(0.07439378471654645,0.02465)
(0.0685228759970794,0.14528)
};
\addlegendentry{MD learning}
\addplot +[thick, color=mycolor!70!black, solid, mark=o, mark options={scale=0.9,solid }] 
coordinates { 
(0.9754782757831633,0.001479)
(0.6805570694630566,0.001537)
(0.60641198352962,0.0015)
(0.5898012234055263,0.001489)
(0.6576671453918237,0.001526)
(0.25712988283032234,0.001809)
(0.18418023947381934,0.002438)
(0.16834017312400895,0.002704)
(0.15456910456432715,0.00356)
(0.1366900600720249,0.004725)
(0.12521277048100887,0.005826)
(0.1326812073447956,0.011095)
(0.11998450420786055,0.751768)
};
\addlegendentry{NN learning, limited complexity}
\addplot +[thick, color=brown!80!black, solid, mark=o, mark options={scale=0.8,solid }] 
coordinates { 
(0.8539545505025266,0.00149)
(0.5255907568051478,0.00169)
(0.37024977450912366,0.00199)
(0.27070055230929324,0.0019)
(0.2196949882516065,0.00213)
(0.17418997530821878,0.00269)
(0.14536707036332697,0.00322)
(0.12352717800246626,0.00382)
(0.10908142938717902,0.00451)
(0.09825260248860279,0.00603)
(0.08551220565306039,0.00904)
(0.07796289778726395,0.01953)
(0.07019443950268744,0.1261)
};
\addlegendentry{MD learning, limited complexity}

\end{axis}

\end{tikzpicture}

}
            \vspace{-4mm}
            \subfloat[]{\begin{tikzpicture}[font=\scriptsize]
\begin{axis}[
width=8cm,
height=4.cm,
legend cell align={left},
legend columns = 2,
legend style={
  fill opacity=1,
  draw opacity=1,
  text opacity=1,
  at={(1,1)},
  anchor=north east
},
log basis y={10},
tick align=outside,
tick pos=left,
x grid style={white!69.0196078431373!black},
xlabel={RMSE [deg]},
xmajorgrids,
xmin=0, xmax=6,
xtick style={color=black},
y grid style={white!69.0196078431373!black},
ylabel={SER},
ymajorgrids,
ymin=1e-03, ymax=1,
ymode=log,
ytick style={color=black},
]
\addplot+[only marks, mark=triangle*, mark options={scale=0.8,color=red}]
coordinates { 
(5.6916003114126035,0.00296)
(5.5677156296591495,0.00283)
(5.6571216946553715,0.0028)
(5.800283398609363,0.00258)
(5.690496974624519,0.00272)
(5.727713486171062,0.00287)
(5.5761694687103285,0.0028)
(5.679408999678488,0.00271)
(2.1004765427219283,0.00385)
(2.1119505953649336,0.00334)
(2.0893563714575656,0.00316)
(2.107938586948127,0.0028)
(2.1508585490086984,0.00305)
(2.2028290976660023,0.00296)
(2.165616401941411,0.0033)
(2.2062893661261462,0.00355)
(2.0400398353340616,0.00459)
(2.036220440279001,0.00455)
(2.0072447537269715,0.00346)
(2.0141810181964326,0.00331)
(2.020068064794682,0.00345)
(2.0409761840434384,0.00363)
(2.082221626635284,0.00389)
(2.062693020533096,0.00416)
(1.982761300602466,0.00735)
(1.9991463612664617,0.00526)
(1.9578900179734169,0.00432)
(1.9673405846954324,0.00411)
(1.9809162292283096,0.00376)
(1.963199974338069,0.00436)
(1.9931438377523265,0.00502)
(2.008185015216435,0.00578)
(1.9718927556140635,0.01136)
(1.9641316716947423,0.00674)
(1.9356593723841136,0.00558)
(1.954613541691669,0.00473)
(1.9518956606007432,0.00445)
(1.9442472509771345,0.00515)
(1.9744984837683615,0.00584)
(1.9549570509635426,0.00722)
(1.9418905627757093,0.02192)
(1.928569003520566,0.01047)
(1.925964949147067,0.00726)
(1.9213213002035159,0.00682)
(1.9319074519414954,0.0066)
(1.932847585174951,0.0077)
(1.9291642900794999,0.00912)
(1.9494799162880025,0.01121)
(1.9360884485403622,0.04755)
(1.943001233494801,0.01647)
(1.9214265533016297,0.01123)
(1.9172817715750177,0.01161)
(1.91766488382922,0.01299)
(1.9069261097373789,0.01522)
(1.9082228150194134,0.0178)
(1.9198665641368888,0.02384)
(1.9019053711915994,0.09174)
(1.9067311287438948,0.09311)
(1.9086120273855562,0.09068)
(1.9145951622360504,0.09205)
(1.8994876434526924,0.0904)
(1.918353052782741,0.09178)
(1.9001703042235982,0.09241)
(1.8970358063361112,0.09226)
};
\addplot +[thick, color=blue!50!black, solid, mark=*, mark options={scale=0.8,solid }] 
coordinates { 
(27.10730830761387,0.001366)
(3.0533789337083497,0.001367)
(1.5612364155483482,0.001473)
(1.425453106611801,0.001447)
(1.5776499575574672,0.001445)
(0.948356109779309,0.001676)
(0.7906262837647577,0.002231)
(0.7542649019497426,0.002624)
(0.7406646052389536,0.003079)
(0.7097890639114209,0.004664)
(0.6982142866695996,0.006301)
(0.6904228715192562,0.007911)
(0.7338812446719374,0.749921)
};
\addplot +[thick, color=green!60!black, solid, mark=diamond*, mark options={scale=0.8,solid}]
coordinates { 
(2.8218291303491756,0.00159)
(1.4691653431959513,0.00153)
(1.2550567422134016,0.00187)
(1.1450576240531833,0.00195)
(1.033297002568827,0.00227)
(0.9615529237561513,0.00251)
(0.9296075233671287,0.00346)
(0.8948520596717418,0.00383)
(0.8780012062829753,0.00495)
(0.8284342183425139,0.00716)
(0.8273092707255159,0.01127)
(0.806102612587278,0.02465)
(0.7952208162311428,0.14528)
};
\addplot +[thick, color=mycolor!70!black, solid, mark=o, mark options={scale=0.9,solid }] 
coordinates { 
(35.59378604216337,0.001479)
(2.673817010410198,0.001537)
(3.2178201726194344,0.0015)
(2.50734842616774,0.001489)
(4.8710216007802565,0.001526)
(3.666406909050489,0.001809)
(2.743188640434559,0.002438)
(2.9508501209314333,0.002704)
(2.5512414155110514,0.00356)
(2.695763004800289,0.004725)
(1.946353693269884,0.005826)
(1.6660776155157233,0.011095)
(1.7561843882056074,0.751768)
};
\addplot +[thick, color=brown!80!black, solid, mark=o, mark options={scale=0.8,solid}]
coordinates { 
(2.493038527199614,0.00149)
(1.397236787711145,0.00169)
(1.2018297564303357,0.00199)
(1.1026156694821918,0.0019)
(1.0062032002522725,0.00213)
(0.9484640988777407,0.00269)
(0.8935985332849942,0.00322)
(0.85680614783988,0.00382)
(0.8526909926538989,0.00451)
(0.8250298138618781,0.00603)
(0.8099722990585945,0.00904)
(0.7741955334013643,0.01953)
(0.776374707927818,0.1261)
};

\end{axis}

\end{tikzpicture}

}
            \vspace{-4mm}
            \hspace*{\fill} 
            \caption{Results under hardware impairments. The target lies in the angular sector $[\thetamin, \thetamax] = [-40\degree, -20\degree]$. The communication receiver is randomly located in the angle interval $[\thetamin, \thetamax] = [30\degree, 50\degree]$. The \ac{ISAC} scenario has fixed testing parameters $\pfa = 10^{-2}$, $\SNR_c = 20\dB$, and $\SNR_r = 0\dB$.}
            \label{fig:ser_vs_pd_hwi_1} \vspace{-4mm}
        \end{figure}

        We now consider hardware impairments, which consist of perturbing the spacing between antenna elements in the transmitter \ac{ULA} as $d_k \sim_{\text{i.i.d.}} \mathcal{N}(\lambda/2, \sigma_{\lambda}^2)$, with $k=0,\ldots, K-2$. We assume $\sigma_{\lambda} = \lambda/30$. Fig.~\ref{fig:ser_vs_pd_hwi_1} shows the \ac{ISAC} trade-off curves for a single realization of $d_k$. The main difference with respect to Fig.~\ref{fig:ser_vs_pd_1} is that the performance of the baseline drops in terms of angle estimation. This occurs naturally when the assumed models differ from reality (the assumed steering vector differs for hardware impairments). Conversely, when complexity is not limited, both end-to-end learning approaches are able to adapt to the impairments and show good performance, although neural-network learning slightly outperforms model-driven learning.
        
        For complexity limitation, the number of hidden neurons in the neural-network-based learning is reduced to $N=21$, and the grid points in $\bmA$ to $\Ngrid = 156$. This makes the number of trainable parameters of the sensing neural networks (approximately 5,000 real-valued parameters) comparable to the model-driven approach (approximately 2,500 complex-valued parameters). Neural-network-based degrades both for detection probability and angle estimation when the number of parameters is reduced. However, model-driven learning shows similar performance with respect to unlimited complexity. This indicates that in cases where complexity is limited, model-driven approaches perform better than neural-network-based learning.
        
    \subsection{Generalization Results}
    \label{subsec:generalization}
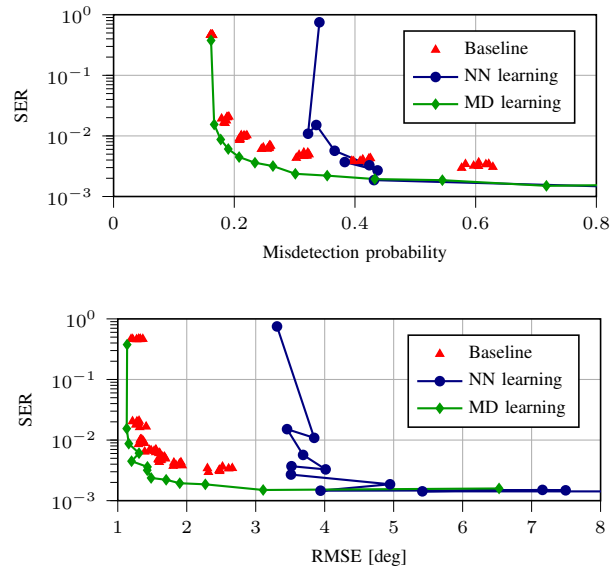
\begin{figure}[t]
        \captionsetup[subfigure]{labelformat=empty}
            \centering
            \subfloat[]{\begin{tikzpicture}[font=\scriptsize]
\begin{axis}[
width=8cm,
height=4.cm,
legend cell align={left},
legend style={
  fill opacity=1,
  draw opacity=1,
  text opacity=1,
  at={(0.6,0.4)},
  anchor=south west
},
log basis y={10},
tick align=outside,
tick pos=left,
x grid style={white!69.0196078431373!black},
xlabel={Misdetection probability},
xmajorgrids,
xmin=0, xmax=0.8,
xtick style={color=black},
y grid style={white!69.0196078431373!black},
ylabel={SER},
ymajorgrids,
ymin=1e-03, ymax=1,
ymode=log,
ytick style={color=black},
]
\addplot+[only marks, mark=triangle*, mark options={scale=0.8,color=red}]
coordinates { 
(0.984258740349614,0.00256)
(0.9847057313903985,0.00261)
(0.9848937026025714,0.00261)
(0.9842810451996487,0.00321)
(0.9845769817516973,0.00266)
(0.9840948848339814,0.00267)
(0.9832089599674553,0.00271)
(0.9844803929014455,0.00253)
(0.604577650491011,0.00318)
(0.6216727065275878,0.00344)
(0.6285054804384351,0.00307)
(0.6168941120485947,0.00339)
(0.6049799196787149,0.00363)
(0.5966596055286321,0.00321)
(0.5831003532064455,0.00342)
(0.5767210535852538,0.00296)
(0.4214477652537535,0.00374)
(0.4256152963467029,0.00428)
(0.4231936814668732,0.00423)
(0.4220214000983231,0.00395)
(0.412883066581306,0.00407)
(0.4085671972919811,0.00385)
(0.39753718048898135,0.00381)
(0.3947502316883461,0.00386)
(0.3191307345295529,0.00471)
(0.3239521506093058,0.00494)
(0.3218425032594524,0.00528)
(0.32090885174459827,0.00473)
(0.31518981898210896,0.00464)
(0.3152531376302915,0.00529)
(0.30741950407105845,0.00486)
(0.3027894327894328,0.00433)
(0.2578193102393198,0.00628)
(0.25859190553128686,0.00696)
(0.2583905669006382,0.00673)
(0.25772564738450887,0.00661)
(0.26011421711383786,0.00686)
(0.254782263705954,0.00611)
(0.24499270325281308,0.00607)
(0.24812602951670237,0.00642)
(0.21530625261605307,0.00961)
(0.2208556137016019,0.0102)
(0.21618827537786234,0.01014)
(0.21158327917208286,0.01023)
(0.21800578988876507,0.00983)
(0.20960456031252483,0.00891)
(0.20888985874103494,0.00854)
(0.20853178819896945,0.0091)
(0.1857635353564081,0.0182)
(0.18869792768513316,0.02026)
(0.1910073684323298,0.02073)
(0.18771031436959185,0.02035)
(0.17898741618317238,0.01927)
(0.18681348397569475,0.01793)
(0.1826481397983989,0.01664)
(0.1846050422687211,0.01627)
(0.16156393251890633,0.46208)
(0.16386113562245574,0.45893)
(0.16112631558004897,0.46536)
(0.16050161690995712,0.46242)
(0.16327158021229726,0.46432)
(0.16402220764617692,0.46399)
(0.16355242261058722,0.46383)
(0.16115796977360064,0.46522)
};\addlegendentry{Baseline};
\addplot +[thick, color=blue!50!black, solid, mark=*, mark options={scale=0.8,solid }] 
coordinates { 
(0.9862758806870813,0.001417)
(0.9282196277980787,0.001425)
(0.845542974263482,0.001505)
(0.8074076412649082,0.001481)
(0.8183928595607394,0.001461)
(0.4315058892679037,0.001864)
(0.43745000959429314,0.002696)
(0.3830934153036143,0.003703)
(0.42384083584552634,0.00328)
(0.3663233352892281,0.005685)
(0.33587915352370745,0.015142)
(0.3222526632588155,0.010872)
(0.34105234841268117,0.750601)
};
\addlegendentry{NN learning}
\addplot +[thick, color=green!60!black, solid, mark=diamond*, mark options={scale=0.8,solid }] 
coordinates { 
(0.9314647007927414,0.00159)
(0.7172296586353405,0.0015)
(0.5448294037067613,0.00186)
(0.4328962473450224,0.00194)
(0.35397554269011966,0.00221)
(0.3011826583364937,0.00237)
(0.2640902207699052,0.00318)
(0.23426522175235598,0.00361)
(0.20802557159391177,0.00448)
(0.19002052709576644,0.00608)
(0.17754930953302805,0.00872)
(0.16671985886692597,0.01546)
(0.16151817818617653,0.37626)
};
\addlegendentry{MD learning}

\end{axis}

\end{tikzpicture}

}
            \vspace{-4mm}
            \subfloat[]{\begin{tikzpicture}[font=\scriptsize]
\begin{axis}[
width=8cm,
height=4.cm,
legend cell align={left},
legend style={
  fill opacity=1,
  draw opacity=1,
  text opacity=1,
  at={(0.6,0.4)},
  anchor=south west
},
log basis y={10},
tick align=outside,
tick pos=left,
x grid style={white!69.0196078431373!black},
xlabel={RMSE [deg]},
xmajorgrids,
xmin=1, xmax=8,
xtick style={color=black},
y grid style={white!69.0196078431373!black},
ylabel={SER},
ymajorgrids,
ymin=1e-03, ymax=1,
ymode=log,
ytick style={color=black},
]
\addplot+[only marks, mark=triangle*, mark options={scale=0.8,color=red}]
coordinates { 
(14.260293545750857,0.00256)
(13.965496505981319,0.00261)
(13.671928214201783,0.00261)
(13.282998744123354,0.00321)
(13.510553558971141,0.00266)
(14.49629162727484,0.00267)
(14.487093243466823,0.00271)
(14.399663877141684,0.00253)
(2.4696690668721994,0.00318)
(2.6606731587393093,0.00344)
(2.476555399912842,0.00307)
(2.606719340922688,0.00339)
(2.518549832434747,0.00363)
(2.47703506768166,0.00321)
(2.2998527503707487,0.00342)
(2.3141883559938035,0.00296)
(1.8039535311906036,0.00374)
(1.9159466403600038,0.00428)
(1.8132984951883315,0.00423)
(1.8218077213000041,0.00395)
(1.8947859033941077,0.00407)
(1.8438471370334117,0.00385)
(1.9339167257930705,0.00381)
(1.8181942911213338,0.00386)
(1.5916985448769458,0.00471)
(1.691940668039607,0.00494)
(1.6520744782867844,0.00528)
(1.6321846920497727,0.00473)
(1.6199645931132056,0.00464)
(1.677472996462143,0.00529)
(1.5869159614375765,0.00486)
(1.60147443703912,0.00433)
(1.549826616282091,0.00628)
(1.5524220744356667,0.00696)
(1.484242295680877,0.00673)
(1.5265030156353543,0.00661)
(1.4612567947728325,0.00686)
(1.6078822066649554,0.00611)
(1.5964121169461303,0.00607)
(1.3902931578602107,0.00642)
(1.3698552281107557,0.00961)
(1.3472945613399672,0.0102)
(1.3277755212767661,0.01014)
(1.3318068743722244,0.01023)
(1.3406598065493673,0.00983)
(1.386895029562493,0.00891)
(1.3014262264694922,0.00854)
(1.3482615016185595,0.0091)
(1.2821703366713204,0.0182)
(1.277724841144159,0.02026)
(1.3084319129216038,0.02073)
(1.2136006840263898,0.02035)
(1.3052024699786915,0.01927)
(1.2629990647409892,0.01793)
(1.4112615104612058,0.01664)
(1.3170349750513821,0.01627)
(1.323818660950487,0.46208)
(1.2726230959783413,0.45893)
(1.3630218978882367,0.46536)
(1.3009144134243358,0.46242)
(1.328073866864411,0.46432)
(1.215460481155003,0.46399)
(1.1924076833416697,0.46383)
(1.3037603930757227,0.46522)
};\addlegendentry{Baseline};
\addplot +[thick, color=blue!50!black, solid, mark=*, mark options={scale=0.8,solid }] 
coordinates { 
(11.37192261398126,0.001417)
(5.416011989361534,0.001425)
(7.157595212291074,0.001505)
(7.493654814021191,0.001481)
(3.93960035218645,0.001461)
(4.947308596717238,0.001864)
(3.511074573560164,0.002696)
(3.5184603751248464,0.003703)
(4.013745604476917,0.00328)
(3.686898424692078,0.005685)
(3.4542061583766865,0.015142)
(3.847971535070118,0.010872)
(3.30876208044973,0.750601)
};
\addlegendentry{NN learning};
\addplot +[thick, color=green!60!black, solid, mark=diamond*, mark options={scale=0.8,solid}]
coordinates { 
(6.5282868061320345,0.00159)
(3.1078999006684773,0.0015)
(2.269377591614702,0.00186)
(1.8987649881460593,0.00194)
(1.7013454221219717,0.00221)
(1.483177018839165,0.00237)
(1.4268709351896098,0.00318)
(1.430030171179367,0.00361)
(1.1999342284011785,0.00448)
(1.3079417022073143,0.00608)
(1.1592159351009599,0.00872)
(1.129723034546281,0.01546)
(1.13539730811437,0.37626)
};
\addlegendentry{MD learning};

\end{axis}

\end{tikzpicture}

}
            \vspace{-4mm}
            \caption{Results under hardware impairments  with low complexity constraints. The target lies in an angular sector $[\thetamin, \thetamax] = [-20\degree, 20\degree]$ which is not included in the training dataset. The communication receiver is randomly located in the angle interval $[\thetamin, \thetamax] = [30\degree, 50\degree]$. The \ac{ISAC} scenario has fixed testing parameters $\pfa = 10^{-2}$, $\SNR_c = 20\dB$, and $\SNR_r = 0\dB$.}
            \label{fig:ser_vs_pd_hwi_gen} \vspace{-4mm}
        \end{figure}

        We now assess the generalization performance of the considered schemes via the testing scenario $[\thetamin, \thetamax] = [-20\degree, 20\degree]$ which is not included in the training dataset.
        Fig.~\ref{fig:ser_vs_pd_hwi_gen} depicts the results assuming no complexity restrictions. We expected that the baseline would outperform learning approaches since they are tested on new unseen data. However, model-driven learning is the best approach, outperforming neural-network learning for both target detection and angle estimation. This implies that the model-driven approach does not overfit to the training data, and it also captures the model structure of the impaired steering vectors.
        

\section{Conclusions}
\label{sec:conclusions}
In this work, we have developed a model-driven ML approach for  \ac{ISAC} and compared to both a neural-network-based ML approach and a model-based baseline. 
Under hardware imperfections in the transmitter \ac{ULA}, both learning methods outperform the model-based baseline since the assumed model differs from reality. 
In addition, the model-driven learning approach outperforms neural-network-based learning under complexity constraints and shows better generalization behavior for testing scenarios that are not seen during training. 
%
In future works, complexity reduction can be carried out applying pruning \cite{reed1993pruning} techniques to the neural networks. In addition, the sample complexity of the proposed approach could be optimized, for example by introducing physically motivated constraints on the weight matrix $\mathbf{A}$, as in \cite{Chatelier2022}. Moreover, the time complexity of training could potentially be reduced using hard thresholding that produces sparse activations in the network, as in \cite{Lemagoarou2021}.
\section*{Acknowledgment}
{
{The authors  gratefully acknowledge the feedback from Juliano Pinto and Lennart Svensson.}}


\balance

\bibliographystyle{IEEEtran}
\bibliography{ref}

\end{document}